# Branch Mode Selection during Early Lung Development

Denis Menshykau[1], Conradin Kraemer[1], Dagmar Iber[1,2]*

1 Department for Biosystems Science and Engineering, ETH Zurich, Basel, Switzerland, 2 SIB, Basel, Switzerland

**Abstract**

Many organs of higher organisms, such as the vascular system, lung, kidney, pancreas, liver and glands, are heavily branched structures. The branching process during lung development has been studied in great detail and is remarkably stereotyped. The branched tree is generated by the sequential, non-random use of three geometrically simple modes of branching (domain branching, planar and orthogonal bifurcation). While many regulatory components and local interactions have been defined an integrated understanding of the regulatory network that controls the branching process is lacking. We have developed a deterministic, spatio-temporal differential-equation based model of the core signaling network that governs lung branching morphogenesis. The model focuses on the two key signaling factors that have been identified in experiments, fibroblast growth factor (FGF10) and sonic hedgehog (SHH) as well as the SHH receptor patched (Ptc). We show that the reported biochemical interactions give rise to a Schnakenberg-type Turing patterning mechanisms that allows us to reproduce experimental observations in wildtype and mutant mice. The kinetic parameters as well as the domain shape are based on experimental data where available. The developed model is robust to small absolute and large relative changes in the parameter values. At the same time there is a strong regulatory potential in that the switching between branching modes can be achieved by targeted changes in the parameter values. We note that the sequence of different branching events may also be the result of different growth speeds: fast growth triggers lateral branching while slow growth favours bifurcations in our model. We conclude that the FGF10-SHH-Ptc1 module is sufficient to generate pattern that correspond to the observed branching modes.

**Funding:** DM was funded through an ETH Fellowship. The funders had no role in study design, data collection and analysis, decision to publish, or preparation of the manuscript.

**Competing Interests:** The authors have declared that no competing interests exist.

* E-mail: dagmar.iber@bsse.ethz.ch

## Introduction

Branched structures are ubiquitous in nature, and the mechanism of their formation has been investigated for decades both in experimental [1–3] and theoretical studies [4,5]. Studies some 50 years ago showed that the dimension of the airway in adult lungs depends exponentially on the branch order very well up to the 10th generation [6]; based on subsequent analysis lungs were suggested to be fractals, with fractal dimension close to 3 for the adult human lung [7–14]. Various algorithms that generate lung trees with morphometric characteristics similar to adult mammalian lungs have been reported [4,5,15,16]. In a geometrically realistic model Kitaoka and co-workers required nine basic and four complementary rules to fill the 3-dimensional thoracic cavity with a branching model [16]. The rules that they defined were to a large part similar to those that were later discovered in a careful experimental study of the growing lung [17]. In this later experimental study it was shown that the branching process during lung development is remarkably stereotyped and that the branched tree is generated by the sequential use of three geometrically simple modes of branching, i.e. domain branching, planar bifurcation, and orthogonal bifurcation [17]. Errors like branch displacement are observed in less than 1% of all branching events [17]. Branching is thus not a stochastic process, but must be controlled. This raises the question of how the information required to generate a structure of such complexity is encoded in the genome.

Since branched structures are created by recursive processes, a limited number of proteins can, in principle, control all of the branching in the lung from the trachea to the terminal bronchioles [3]. Genetic studies have led to the identification of key regulatory genes and morphogenes that control lung development, most importantly Fibroplast growth factor (FGF)10, Sonic hedgehog (SHH), and Bone morphogen protein (BMP)4 [1,2,18–23] (Figure 1a). FGF10 is expressed in the distal mesenchyme around the epithelial bud tip and is essential for the lung bud formation, proliferation of the endoderm, and directional outgrowth [3,24,25]. FGF10 signals mainly through the epithelial FGF receptor FGFR2, and inactivation of FGFR2 in the lung epithelium results in the disruption of lobes and small epithelial outgrowths that arise arbitrarily along the main bronchi [25]. FGF induces the expression of SHH, BMPs, and Sprouty which in turn limit the expression and signalling of FGFs [24,26–29]. SHH signals through its receptor Patched (Ptc), and affects FGF10 and BMP4 activity. While many studies show that BMPs regulate lung development their detailed effects have been difficult to disentangle [1]. Adding BMP4 to organ cultures of the whole embryonic lung promotes branching morphogenesis and increases the number of peripheral epithelial buds [30]. Addition of exogenous BMP4 protein to mesenchyme-free endoderm cultured *in vitro* with FGFs, on the other hand, inhibits proliferation, secondary budding and differentiation [28,29,31]. Both overexpression of BMP4 and a conditional knock-out result in similar lung phenotypes [32,33], suggesting that correct BMP4 levels are essential for normal lung

### Author Summary

Most organs of higher organisms, such as the vascular system, lung, kidney, pancreas, liver and glands, are heavily branched structures. The branching process during lung development has been studied in great detail and is remarkably stereotyped. The branched tree is generated by the sequential, non-random use of three geometrically simple modes of branching. While the branching sequence is identical in mice of identical genetic background it differs between mouse strains. This suggests that the positioning of branch points and the type of branching sensitively depends on information encoded in the genome. Encoding every branching point independently in the genome would require a large number of genes, and it is more likely that a recursive, self-organized process exists that determines the patterning. While many regulatory molecules have been identified an integrated understanding of the regulatory network (program) is missing. Based on available experimental data we have developed a model for lung branching. The model correctly predicts branching phenotypes in mutants and suggests that also the growth speed of the lung tip can affect the positioning and type of the next branching event.

development. These may be maintained by the many negative feedbacks that control BMP activity.

Theoretical studies suggest that physical forces can play a key role in lung branching morphogenesis [34–37]. Recent studies demonstrate that the mechanical stresses do influence branching morphogenesis [38]. Increased internal pressure leads to an increase of lung branching in in vitro cultures [39], and cellular contractility is critical for branching morphogenesis of the lung. Inhibiting actomyosin-mediated contractility in whole lung explants decreases branching [40], whereas activating contractility increases branching [41]. A qualitative model that described epithelial branching in culture experiments showed that both the mechanical strength of the cytoskeleton and the reaction-diffusion kinetics can in principle affect branching morphogenesis [36,42]. Work on mammary glands further suggests that the geometry of tubules might dictate the position of branches [43].

Computational models can explore the impact of the signaling interactions, physical forces and domain geometries and thus discern a minimal set of rules and interactions from which the observed pattern can emerge. Hirashima and co-workers recently proposed a simple three component model on a 2-dimensional lung bud cross-section [44] to explain the mechanistic basis of different branching modes. The model focused on the interactions between SHH, transforming growth factor(TGF)-beta and FGF10 and suggests that domain length and shape can have a strong impact on the distribution of morphogenes and the selection of branching points in the developing lung. This prediction, however, hinges on a particular distribution of TGF-beta (constant at the stalk) and SHH (fixed at the tip) and is valid only with a particular type of boundary condition (impermeable lung boundaries) which is unlikely to apply (Figure S1).

Turing-type models have been suggested as an alternative to explain the emergence of regular patterns as observed during lung branching morphogenesis [45,46]. However, to our knowledge no such mechanistic Turing model has yet been formulated for the lung. Here we present a reaction-diffusion model that we developed based on available information from the literature. The Turing-type model reproduces available experimental data both in wild type and mutant mice and provides a mechanistic explanation for the different lung branching types. We further show that the rate with which the lung bud grows can determine the branching mode.

## Model

We sought to develop a model for the core signaling module that regulates branch point selection during lung development. Many proteins have been implicated in the branching process, but FGF10 and SHH appear to play the most prominent roles [1,19]. Culture experiments have identified FGF10 as one of the key regulators of lung branching and outgrowth [20,29,47]. Another important protein is SHH which signals via its receptor PTC [48,49]. In spite of its importance we will not include BMP4 in this model [30,50]. This is because BMP4 and SHH both have similar impacts on FGF10 in that both are positively regulated by FGF10 while they negatively affect FGF10 upon binding to their receptors. BMP4, however, cannot directly feed back onto Fgf10 expression because its main receptor, Alk3 [51], is expressed only in the epithelium while FGF10 is expressed in the mesenchyme [1]. Any direct negative impact of BMP4 on FGF10 signaling must thus arise from its interference with FGF10-dependent signaling in the epithelium rather than with Fgf10 expression. As such BMP4 signaling may reduce the extent to which FGF10 promotes Shh expression or may affect other effectors that impact on Fgf10 expression. Since we are interested only in the core patterning mechanism we will ignore the modulating impact of BMP4 in this work, and focus on FGF10, SHH and its receptor Patched (Figure 1a).

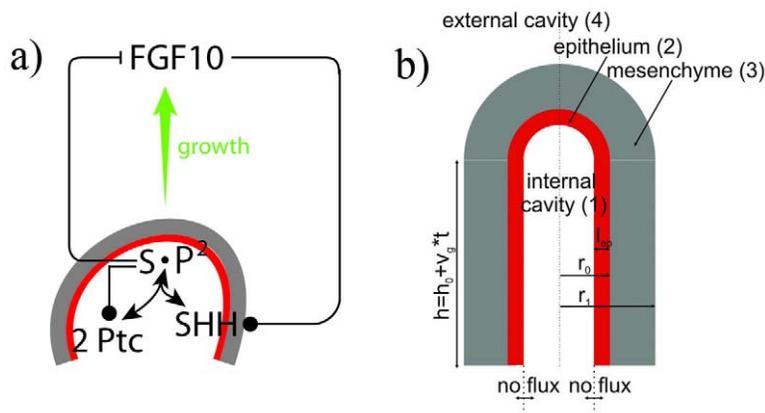

**Figure 1. A graphical summary of the modelled interactions of the signaling factors in lung bud during morphogenesis. a**) FGF10 is transcribed at high levels in the distal mesenchyme (grey) and experiments suggest that FGF10 promotes both the proliferation of the endoderm and its outward movement (green arrow). FGF10 stimulates the expression of SHH in the epithelium (red). SHH reversibly binds its receptor Ptc1 which is expressed in the mesenchyme (grey). SHH-Ptc binding results in the repression of FGF10 expression. **b**) The idealized computational domain comprises a 2D crossection along the cylinder axis of symmetry. The epithelium and the mesenchyme are shown in red and grey, correspondingly. SHH and FGF10 (but not Ptc) can diffuse freely ($D_{ext}$) in the interstitial space (4) and lumen (1). The time-dependent height of the cylinder is $h(t) = h_0 + v_g \times t$.
doi:10.1371/journal.pcbi.1002377.g001

The simulated cross-section of the lung tip contains about 10–20 epithelial cells (Figure 1b). Previous studies have successfully described the *in vivo* distribution of morphogens with continuous reaction-diffusion equations on a domain containing around 10 cells [52–54]; we therefore expect that continuous reaction-diffusion equations are also adequate in our study of the lung tip cross-section. Both FGF (which we denote by $F$) and SHH (which we denote by $S$) can diffuse rapidly [52–54] and we write $\overline{D}_F$ and $\overline{D}_S$ for the diffusion coefficients. Patched-1 receptors (denoted by $P$) are membrane proteins and are therefore restricted to the surface of single cells where they diffuse with a much reduced diffusion coefficient $\overline{D}_P \ll \overline{D}_S, \overline{D}_F$ [55,56]. The exact details of this restricted diffusion appear, however, not to qualitatively affect the Turing pattern since we obtain qualitatively similar Turing pattern in a 2D continuous plane and when we solve the Turing model on a domain with an array of cells where receptors can diffuse only on the surface of the cells and the ligand only in the volume between the cells (Figure S2). Similarly simulations in which the computational domain is split into domains of cell size with Ptc spatially restricted to each cell give qualitatively similar results while being computationally much more costly (Figure S3). This may be accountable to the fact that during the receptor half-life time ($t_{1/2} = \ln(2)/\delta_P = 670\ s$) a receptor diffuses only roughly the distance of one cell diameter ($l = (2Dt_{1/2})^{0.5} \simeq 10\ \mu m$). In conclusion a mean-field approximation with a small, non-zero value for the receptor diffusion coefficient appears to provide a good and computationally efficient approximation. Finally, we note that receptors can move also passively with migrating and dividing cells such that the effective physiological Ptc diffusion coefficient is larger than the diffusion coefficient of Ptc within the cell membrane. We will write $\overline{D}\overline{\Delta}[\cdot]$ for the diffusion fluxes where $\overline{\Delta}$ denotes the Laplacian operator in Cartesian coordinates, and $[\cdot]$ concentration. The characteristic length of gradients depends both on the speed of diffusion and the rate of morphogen removal. In the absence of contrary experimental evidence we will assume the simplest relation, linear decay, at rates $\overline{\delta}_k[k]$ for all components (i.e. $k = F, S, P, C$).

FGF10, SHH and Ptc-1 regulate each other as graphically summarized in Figure 1a. Thus FGF10 expression is repressed by a complex of SHH bound to Ptc-1. We describe the inhibition of FGF10 production by the SHH-Ptc complex, $C$, by a Hill-type function, $\overline{\rho}_F \frac{\overline{K}_F^n}{[C]^n + \overline{K}_F^n}$, with Hill constant $K_F$, Hill coefficient $n$, and $\overline{\rho}_F$ as the maximal rate of FGF10 production in the absence of SHH and obtain the following equation describing the spatio-temporal dynamics of the FGF10 concentration

$$[\dot{F}] = \underbrace{\overline{D}_F \overline{\Delta}[F]}_{diffusion} + \underbrace{\overline{\rho}_F \frac{\overline{K}_F^n}{[C]^n + \overline{K}_F^n}}_{production} \underbrace{-\overline{\delta}_F[F]}_{degradation}, \quad (1)$$

where we use $[\dot{X}] = \frac{\partial [X]}{\partial t}$ as short-hand notation for the time derivative.

FGF10 stimulates the expression of SHH and the rate of SHH expression is therefore best described by a Hill-type function, $\overline{\rho}_S \frac{[F]^n}{[F]^n + \overline{K}_S^n}$ with Hill constant $\overline{K}_S$ and Hill coefficient $n$. SHH is a multimer and one SHH molecule can therefore bind at least two receptors [57]. We therefore use as rate of complex formation $k_{on}[S][P]^2$ and $k_{off}[C]$ as rate of dissociation and obtain for the SHH dynamics:

$$[\dot{S}] = \underbrace{\overline{D}_S \overline{\Delta}[S]}_{diffusion} + \underbrace{\overline{\rho}_S \frac{[F]^n}{[F]^n + \overline{K}_S^n}}_{production} \underbrace{-\overline{\delta}_S[S]}_{degradation} \\ \underbrace{-\overline{k}_{on}[P]^2[S] + \overline{k}_{off}[C]}_{complex\ formation} \quad (2)$$

The expression of Ptc-1 is enhanced in response to binding of SHH to the Ptc-1 receptor, and the rate of Ptc-1 expression must therefore be a function of the concentration of the complex, $C$, i.e. $\mu([C])$. Free Ptc-1 is removed by complex formation and restored by its dissociation such that the spatio-temporal dynamics of Ptc can be described by

$$[\dot{P}] = \underbrace{\overline{D}_P \overline{\Delta}[P]}_{diffusion} + \underbrace{\mu([C])}_{production} \underbrace{-\overline{\delta}_P[P]}_{degradation} \underbrace{-2k_{on}[P]^2[S] + 2\overline{k}_{off}[C]}_{complex\ formation} \quad (3)$$

If we assume that the diffusion of the membrane-bound SHH-Ptc-1 complex is slow compared to its binding and turn-over kinetics then we can neglect the diffusion operator in the equation and write for the dynamics of the complex $C$

$$[\dot{C}] = \underbrace{k_{on}[P]^2[S] - \overline{k}_{off}[C]}_{complex\ formation} \underbrace{-\overline{\delta}_C[C]}_{degradation}. \quad (4)$$

If the dynamics of the complex are fast compared to those of the other components then we can introduce a quasi steady state approximation, and obtain for the concentration of bound receptor $[C]_{SS}$:

$$[C]_{SS} = \frac{\overline{k}_{on}}{\overline{k}_{off} + \overline{\delta}_C}[P]^2[S] = \overline{\Gamma}[P]^2[S] \quad (5)$$

where $\overline{\Gamma} = \frac{\overline{k}_{on}}{\overline{k}_{off} + \overline{\delta}_C}$. The concentration of bound receptor, $[C]$, is thus proportional to $[P]^2[S]$. We will further use a linear approximation as the simplest possible relation for the receptor production rate $\mu([C])$, and write $\mu([C]) = \overline{\rho}_P + \overline{v}[C]$, where $\overline{\rho}_P$ and $\overline{v}$ are zero and first order rate constants, respectively.

Our model is then based on the follwoing set of three PDEs of reaction-diffusion type:

$$[\dot{S}] = \overline{D}_S \overline{\Delta}[S] + \overline{\rho}_S \frac{[F]^n}{[F]^n + \overline{K}_S^n} - \overline{\delta}_C \overline{\Gamma}[P]^2[S] - \overline{\delta}_S[S] \quad (6)$$

$$[\dot{P}] = \overline{D}_P \overline{\Delta}[P] + \rho_P + (\overline{v} - 2\overline{\delta}_C)\overline{\Gamma}[P]^2[S] - \overline{\delta}_P[P] \quad (7)$$

$$[\dot{F}] = \overline{D}_F \overline{\Delta}[F] + \overline{\rho}_F \frac{\overline{K}_F^n}{(\overline{\Gamma}[P]^2[S])^n + \overline{K}_F^n} - \overline{\delta}_F \quad (8)$$

We note that if $[F] \gg \overline{K}_S$, $\overline{v} = 3\delta_C$, and $\delta_S = 0$ then equations (6)–(7) reduce to the classical Schnakenberg model [58].

## Computational Domain Geometry, Boundary and Initial Conditions

The lung tip can be represented as a cylinder with an internal radius of $r_0 = 50 \mu m$ and an external radius of $r_1 = 100 \mu m$; the thickness of the lung epithelium can be estimated from microscopy data as $l_{ep} = 5$–$10 \mu m$ [32,59]. To keep the calculations as simple as possible while retaining a realistic geometry we use a 2D slice in Cartesian coordinates along the lung bud axis of symmetry as shown in Figure 1b. SHH is produced in the epithelium [1,2] and FGF and Ptc are produced in the surrounding mesenchyme [1,2]. There is no experimental evidence that the mesenchyme or epithelium are surrounded by any insulating layer and accordingly the boundaries are permeable in our simulations. SHH and FGF can thus diffuse unhindered outside of the epithelium and the mesenchyme as well as in the lumen which is filled with a liquid. The unhindered diffusion of SHH and FGF in the lumen and interstitial space is reflected by the diffusion coefficient $\overline{D}_{ext}$ which is much larger than the diffusion coefficients, $\overline{D}_i$, that apply in the tissue; the receptor Ptc is a membrane protein and therefore cannot diffuse into the cavities. The concentrations of FGF and SHH infinitely far from the mesenchyme are assumed to be zero. We note that the predictions of our model are independent of the boundary conditions and similar results are obtained in simulations with zero-flux boundary conditions at the boundary of the lung tissue (see section **Robustness of the Observed Pattern** for details). We start all our simulations with no species present, i.e. we are setting all concentrations to zero at $\tau = 0$.

## Modelling Domain Growth

We consider two modes of 1-dimensional growth along the lung stalk: (1) uniform growth (stretching) of the domain and (2) local growth at the tip of the lung. To conserve mass (rather than concentration) the reaction-diffusion equations (Eqs 6–8) must be expanded to include the advection and dilution terms [60], i.e.

$$[\dot{X}] = D_x \overline{\Delta}[X] + R([X]) + \nabla u[X] + \nabla[X]u \quad (9)$$

where $u$ denotes the growth speed. For homogeneous growth at rate $\overline{v}_g$ we then have

$$[\dot{X}] = D_x \overline{\Delta}[X] + R([X]) + \frac{\dot{h}}{h}[X] + \frac{\dot{h}}{h}\nabla[X] = \frac{\partial[X]}{\partial t} \\ + \frac{\overline{v}_g}{\overline{h}_0 + \overline{v}_g t}[X] + \frac{\overline{v}_g}{\overline{h}_0 + \overline{v}_g t}\frac{\partial[X]}{\partial \overline{y}} \quad (10)$$

where $\overline{h}_0$ denotes the initial height of the lung tip, $\overline{h}(t) = \overline{h}_0 + \overline{v}_g t$ the height of the lung tip, and $t$ denotes time.

## Parameter Values

The domain size and the time scale of the process are well established [17,24]. The measurement of the *in vivo* kinetic parameter values on the other hand is complicated and typically parameter values are known only from experiments in related systems [52–54]. To reduce the number of unknowns we non-dimensionalize the equations and thereby remove five independent parameters. We use $r_0$, the internal radius of the lung bud, as characteristic length scale of the model and $\overline{D}_F/r_0^2$ as its characteristic time scale. Moreover, we non-dimensionalize the FGF concentration with respect to the Hill constant $\overline{K}_S$, i.e.

$F = [F]/\overline{K}_S$, and the SHH and Ptc-1 concentrations with respect to $c_0$, i.e. $S = [S]/c_0$, $P = [P]/c_0$ with $c_0 = \left(\frac{\overline{K}_F}{\overline{\Gamma}}\right)^{1/3}$. Equations 6–8 can then be rewritten in dimensionless form; the dimensionless parameters and variables are summarised in Table S1.

$$\dot{S} = D_S \Delta S + \rho_S \frac{F^n}{F^n + 1} - \delta_S S - \delta_C P^2 S$$

$$\dot{P} = D_P \Delta P + \rho_P - \delta_P P + (v - 2\delta_C)P^2 S$$

$$\dot{F} = \Delta F + \rho_F \frac{1}{(P^2 S)^n + 1} - \delta_F F \quad (11)$$

It should be noted that the Laplacian is now with respect to the non-dimensional space variables and $\dot{X} = \frac{\partial X}{\partial \tau}$.

Five parameters have been removed and the patterning mechanism no longer depends on absolute values of diffusion and decay constants, but only on the relative diffusion coefficients and the relative decay rates. Similarly, the absolute concentrations do not matter, but only the relative concentrations (as a result from the relative expression and decay rates) relative to the Hill constants and the effective binding constant $\overline{\Gamma}$. The value of the Hill coefficient was set to $n = 2$ to account for possible cooperatively effects; however, the model gives very similar results with other values of $n$ (Figure S4). It should be noted here that the stochiometry of SHH and Ptc complex $P_m S_n$ at which patterning is observed is not limited to the case m = 2, n = 1 that we analyse in this manuscript. Similar patterns are observed as long as $m > 1, n > 1$ (see Figure S5 for details).

## Numerical Solution of PDEs

The PDEs were solved with finite element methods as implemented in COMSOL Multiphysics 4.1. COMSOL Multiphysics is a well-established software package and several studies confirm that COMSOL provides accurate solutions to reaction-diffusion equations both on constant [61] and growing two-dimensional domains [62–64]. Mesh and the time step were refined until further refinement no longer resulted in noticeable improvements as judged by the eye (Figure S6). When simulations were performed on an open domain the bulk solution conditions was implemented at a distance $6\sqrt{D\tau_{max}}$ from the mesenchyme, where $\tau_{max}$ is the maximum time of model evaluation. It was shown that beyond this, the effects of diffusion in not important on the experimental time scale [65].

## Local Stability Analysis

A local stability analysis was performed in the following way: parameter values were taken as indicated in Table 1, further parameters were varied one by one until qualitative change of pattern were observed (different number of FGF, SHH and Ptc spots). Accuracy of parameter value estimation was 10% or higher.

## Robustness to Parameter Variability

The approach to estimate robustness to spatial parameter variability was adapted from ref [66]. Parameter values were assumed to be given by the formula $k = k_0 \times (1 + \xi(x,y))$, where $\xi(x,y)$ is normally distributed random function with a mean value of zero and half width $\theta$. The half width of the distribution was equal for all parameters, except geometrical which were not varied.

Table 1. Values of dimensionless parameters used for simulations.

| parameter | domain 2 (epithelium) | domain 3 (mesenchyme) | domain 1 and 4 (cavities) |
|---|---|---|---|
| $D_S$ | 5 | 5 | 40 |
| $D_F$* | 1 | 1 | 40 |
| $D_P$ | 0.02 | 0.02 | - |
| $\delta_C$ | 1.6 | 1.6 | - |
| $\nu$ | - | 5 | - |
| $\delta_S$ | 0.2 | 0.2 | 0.2 |
| $\delta_P$ | - | 1 | - |
| $\delta_F$ | 5 | 5 | 5 |
| $\rho_S$ | 1600 | - | - |
| $\rho_F$ | - | 3.5 | - |
| $\rho_P$ | - | 0.6 | - |

Domain parameters: $l_{ep} = 0.2$, $r_1 = 2$, $h = 1$. $n = 2$.
*Note that the diffusion coefficient of FGF in epithelium and inner radius of mesenchyme $r_0$ are used to nondimensionalise model.
doi:10.1371/journal.pcbi.1002377.t001

## Results

### Morphogen Distribution in the Steady-State

Since the model is an example of a Schnakenberg Turing-type model [58] we expected to find parameter ranges for which we would observe the emergence of FGF patterns on the lung-shaped domain. Indeed such pattern emerged from spatially uniform initial conditions. We note that the observed distribution and expression pattern correspond overall rather well to experimental observations, but since quantitative data on protein concentration distributions are not available we have to restrict ourselves to a qualitative discussion of distribution patterns. We therefore deliberately left out scale bars for better readability. Dark red colours denote the highest concentrations while dark blue colours denote the lowest concentrations. FGF10 induces SHH expression and SHH expression is highest in the epithelium adjacent to the mesenchyme with high FGF10 concentrations (Figure S7b,e) as also observed in experiments [24,32,48,67]. In situ data seem to suggest that the expression domains of Ptc-1 and Shh coincide [24,68,69] but the spatial resolution of the data may be too low to reveal the closed juxtaposition predicted by the model.

Different parameter ranges resulted in steady-state patterns with FGF10 either centered at the lung bud tip (Figure 2a) or concentrated towards the side of the tip (Figure 2d). Experiments have previously demonstrated that lung tips grow towards regions with a high FGF10 concentration [59] and accordingly FGF10 centered at the lung tip will support further outgrowth of the lung tip while FGF10 centered at the sides will induce lateral outgrowth. We thus suggest that the two types of patterns that we observe are likely to correspond to the different branching modes, i.e. lateral branching would thus correspond to the FGF10 pattern in Figure 2a with FGF10 spots at the tip and at the side, while bifurcation would correspond to the FGF10 pattern in Figure 2d with FGF10 centered only to the side but towards the tip. Our model cannot reproduce domain branching or the difference between planar and orthogonal bifurcations as these events are intrinsically three dimensional. Here we should note that the model generates pattern similar to those discussed above if extended to the third dimension (Figure S8). However, a much wider range of patterns is possible in 3D as will be discussed in detail in a separate manuscript (Menshykau and Iber, in preparation). Going forward we will restrict the presentation to the FGF10 concentration pattern as these appear to mainly guide lung outgrowth.

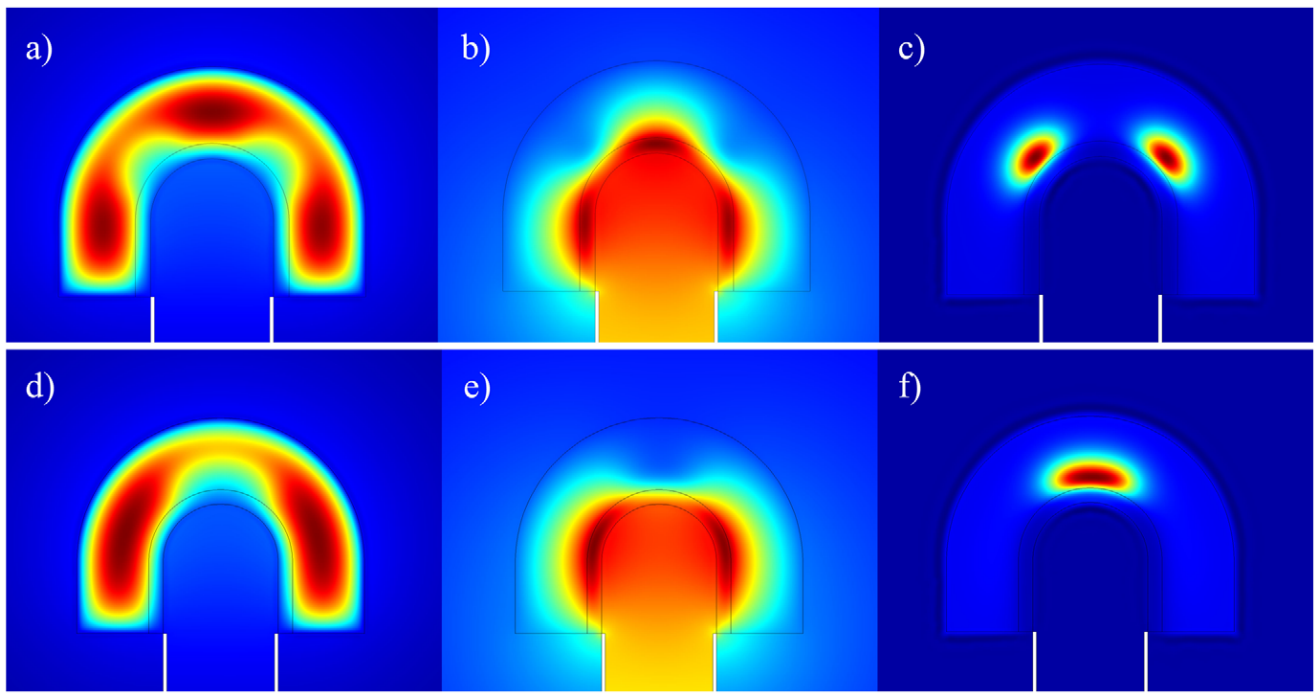

**Figure 2. The steady state distributions of FGF10, SHH and receptor Ptc concentrations.** The steady state distributions of (**a,d**) FGF10, (**b,e**) SHH, and (**c,f**) Ptc for parameter values as in Table 1 (**a–c**) or with $\rho_S = 1300$ (**d–f**). The upper panel presents an example of FGF10 distribution during the lateral branching mode, while the lower panel provides an example for FGF10 distribution during a bifurcation branching mode. Note that the expression patterns of SHH, FGF and Ptc are shown in Figure S7.
doi:10.1371/journal.pcbi.1002377.g002

### Robustness of the Observed Pattern

Since the parameters are difficult to determine accurately in experiments it was important to check how sensitive our results would be to variations in parameter values. We carried out a local stability analysis by altering each parameter value independently. Here we note that the non-dimensional parameters represent relative dimensional parameters. The patterning mechanism is thus robust to a parallel change in the parameter pairs listed in Table S1, and the value of most relative parameters can be changed by 20%–30% without changing the type of the observed pattern (Figure 3). At the same time almost each parameter (except for the degradation rate of the receptor, $\delta_P$) can be employed to switch the pattern between lateral branching (Figure 3, blue) and bifurcation mode (Figure 3, red). The mechanism thus appears to be robust to noise yet sensitive to regulation. We note that virtually all parameter values could be affected by the many additional interactions that have been described but that we chose to ignore in this simple model that focuses only on the core of the regulatory mechanism.

The domain that we chose to solve our model on is an idealization of realistic domains. We therefore also checked the impact of domain deformations (Figure 4). The lateral branching mode is indeed robust to deformations of the domain geometry, i.e. to an increased radius of the epithelium (Figure 4a), the mesenchyme (Figure 4b) or a truncated bud (Figure 4c). Similarly, the boundary conditions are not critical and similar pattern emerge with zero-flux boundary conditions both in the lateral (Figure 4 d) and in the bifurcation mode (Figure 4 e). Linear stability analysis of Equation 11 carried out with parameter values as used to simulate lung branching on the domain with no-flux boundary conditions (Figure 3 d, e) showed that pattern arise because of the diffusion driven instability and therefore Turing type.

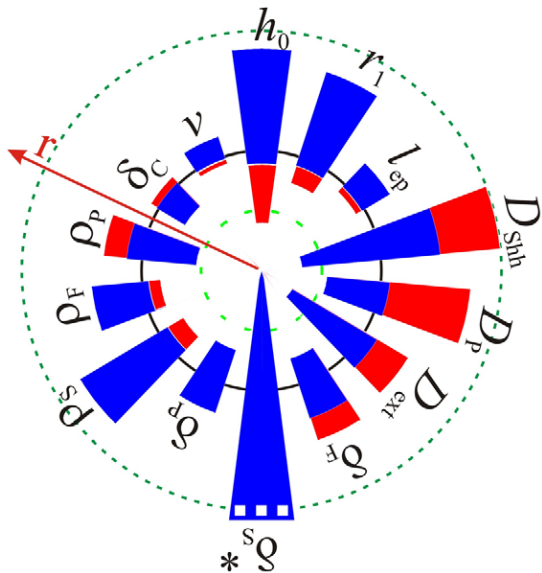

**Figure 3. Local stability analysis.** The blue and red regions represent ranges of the dimensionless parameters for which lateral and planar bifurcation modes of branching are observed respectively. The parameter values in Table 1 are used as a reference point (black solid line), and each parameter was perturbed independently by a factor $r$ as indicated. The green-dashed circles mark halved and doubled parameter ranges. $\delta_S$: * The lateral branching mode (blue) is stable up to value 2.3 times the reference value of $\delta_S$, the bifurcation mode of branching is observed in the range from 2.3 to 7-times the reference value given in Table 1.
doi:10.1371/journal.pcbi.1002377.g003

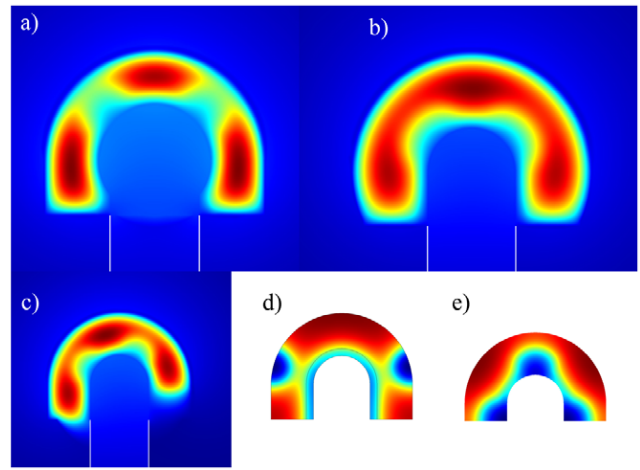

**Figure 4. The FGF10 pattern is robust to changes in the domain geometry and boundary conditions.** The steady state pattern of FGF10 on the computational domain has (**a**) an increased radius of epithelial bud, i.e. $r_0$ at the tip increased by the 25%; (**b**) an increased radius of the mesenchymal bud, i.e. $r_1$ at the tip increased by the 10%; (**c**) a truncated stalk, i.e. $h_0$ at the stalk is truncated by 80%. (**d,e**) The steady state pattern of FGF10 with no flux boundary conditions at the lung boundary: (**d**) all production and degradation rate constant are equal to 0.5 and 1.7 of that presented in Table 1 (lateral branching mode), (**e**) constants are equal to 0.7 and 1.5 of that presented in Table 1 (bifurcation mode). Decrease of production rates and increase of degradation rates are imposed to compensate for the absence of morphogen flux from the epithelium and mesenchyme to the lumen and interstitial space when no-flux boundary conditions are imposed at the lung border. All parameters as in Table 1 unless otherwise stated.
doi:10.1371/journal.pcbi.1002377.g004

### Branching Mode Selection on a Growing Domain

Since lung buds branch as they are growing out we wondered how the growth speed and type would affect patterning. Figure 5 shows the FGF10 distribution on a growing lung bud where growth is restricted to the lung tip. At $v_g = 0.08$ the concentration of FGF is high at the tip of the growing lung, and further regions of high concentration of FGF10 appear at the lung stalk as the lung grows out (Figure 5 a). Importantly new regions of high FGF10 concentration emerge close to the tip. This pattern corresponds to the lateral branching mode. At a 4-fold lower growth speed, $v_g = 0.02$ (Figure 5 b), regions of high FGF concentration appear only at the sides of the lung tip but are absent from the tip itself. This pattern would thus correspond to the bifurcation mode of branching.

If the lung bud grows uniformly in the entire domain the patterning is similar with an important difference: regions with high FGF10 concentration can appear at any position within the domain (Figure S9). Both the insertion of branches directly behind the tip and the insertion of new branches at the proximal side of the stalk have been observed. Similarly, both uniform proliferation (Metzger and co-workers, personal communication), [17] and enhanced proliferation at the tip [70] have been observed. Those experiments that report a concentration of proliferating cells at the tip find 2.5 times more BrdU-stained cells (a marker for dividing cells) at the tip than in the stalk [70], and in our model branching points appear behind the tip as long as the growth rate in the tip is at least 2–3 times larger than in the stalk. We therefore suggest that different lung branches may be growing differently, and that this may explain the different positions at which new branches emerge relative to the tip.

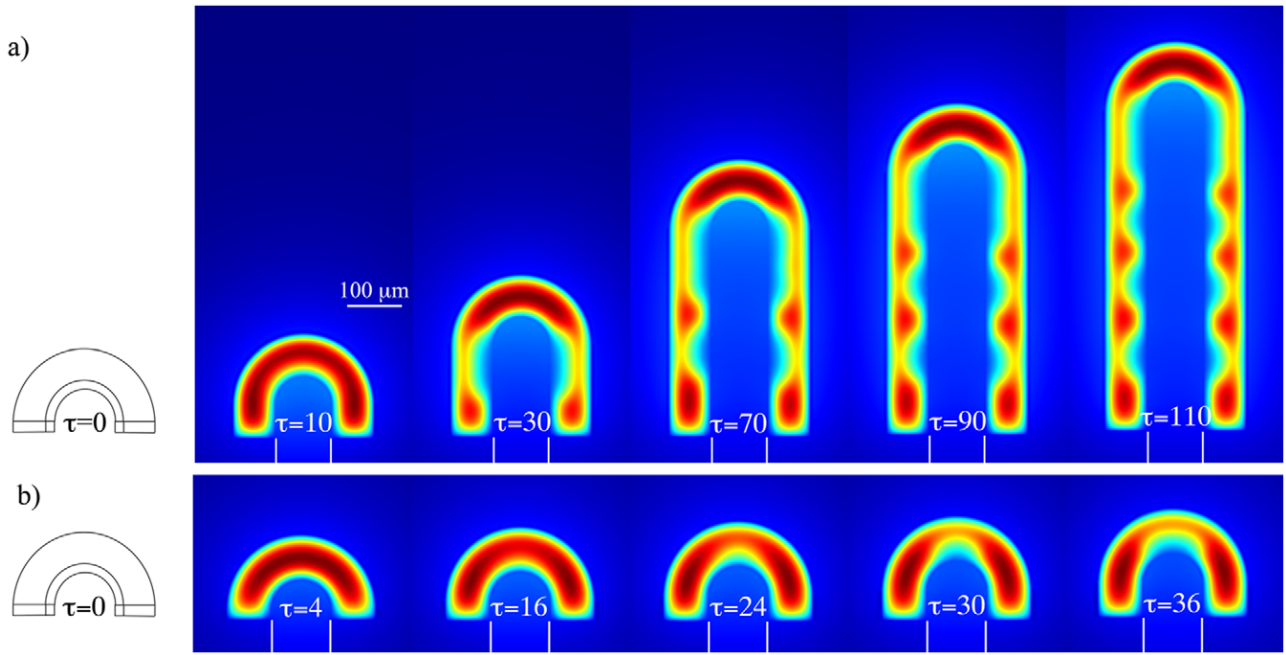

**Figure 5. FGF10 distribution on a growing lung tip domain.** Depending on the growth speed the distribution of FGF10 is either consistent with (**a**) a lateral branching mode (fast growth speed, $v_g = 0.08$), or (**b**) a bifurcating mode of branching (slow growth speed, $v_g = 0.02$). Parameters values used to simulate FGF10 pattern formaton on a growing lung are as given in Table 1, except initial stalk length $h_0 = 0.2$.
doi:10.1371/journal.pcbi.1002377.g005

Earlier we showed that the pattern on a constant domain is robust to small parameter variations. To test that the pattern is robust to small random changes also on growing domains we explored the patterning mechanism if parameter values are drawn from a Gaussian distribution with different standard deviations (see Model section). Figure S10 shows that the domain branching mode remains stable as long as standard deviations of the random variables do not exceed 0.2–0.3 of the reference value presented in Table 1.

### Mutants

An important test for the suitability of a mathematical model is its consistency with a wide range of independent experimental observations. Lung branching morphogenesis has been studied intensively and a large body of experimental results exists to test the model with. These include a large number of in part counterintuitive mutant phenotypes of key signaling proteins in mice [1]. Since our model is restricted to FGF, SHH and Ptc-1 we will focus on mutations in those genes. The computational model reproduces all full knock-outs. However, these results are trivial since the lungs in mutants with fully knocked-out genes are severely truncated or do not form at all [7,49,71–73]. Given the key importance of all three components in the model no pattern is observed if any component is eliminated.

**FGF10.** Experimental evidence suggests that, in addition to its critical roles during lung initiation, FGF10 continues to have important functions also later in lung development. Conditional inactivation of FGF10 in lung mesenchyme resulted in smaller lobes with a reduced number of branches [25]. Interestingly, a 25% reduction in FGF10 expression not only reduces the number of branches but increases the distance between branching points by 50%. [74]. We observe a similar effect in our simulations where a 25% reduction in FGF10 expression ($\rho_F = 2.6$) increases the distance between branching points (Figure 6c). Changes in the FGF10 concentration could, of course, affect the growth speed. However, simulations shows (Figure S11) that at any sensible growth rate mutant lungs have less branching points with an increased distance between them if compared to WT lungs.

FGF10 is expressed only in the mesenchyme, but mesenchyme is not strictly required for lung branching. Thus cultures of isolated endoderm incubated in Matrigel$^{TM}$ substratum still migrate and grow towards an FGF-loaded bead and form branches [28]. This result is in agreement with our model. At constant FGF concentrations [F] as observed in lung cultures the expression of SHH proceeds at contant rate $\rho_S \frac{F^n}{F^n + 1}$ and the patterning module (SHH-Ptc) uncouples from the equation for FGF. As noted above in this limit we recover the classical Schnakenberg Turing model and patterning can still be observed (Figure S12).

**SHH.** SHH plays an important role in lung branching morphogenesis. In Shh null mice the lungs only form a rudimentary sac due to failure of branching and growth after formation of the primary lung buds [49]. Deletion of Shh later in gestation (after E13.5) causes mild abrogation of peripheral branching morphogenesis [75]. Defects in branching morphogenesis and vascularization seen in Shh null mutant (Shh(−/−)) mice can be substantially corrected when SHH is ectopically expressed in the respiratory epithelium [75]. Retinoic-acid enhances the expression of SHH and reduces the expression of FGF10 both by about 60% [76–78] and leads to a decrease in the amount of distal branching. Similarly in the model we predict that strongly enhancing Shh expression and/or strongly reducing FGF10 expression will lead to a loss of pattern (Figure 3). The overexpression of SHH in transgenic mice leads to an upregulation of Ptc while other genes such as BMP4 were not affected [48]. When we reduce (or enhance) SHH production ($\rho_S$) in our simulations then the production of Ptc in a growing lung is heavily delayed (or speeded up respectively) (Figure 6e) which may well explain why Ptch appears upregulated when Shh is overexpressed. Mutants with enhanced diffusion of SHH have not been described in detail but, unlike for the limb, no abnormalities have been reported in lung development [79]. An increase of the SHH diffusional coefficient in the model results in a gradual increase in the distance between branching modes (Figure 6c,d). However, no such effect is visible up to a 50% increase.

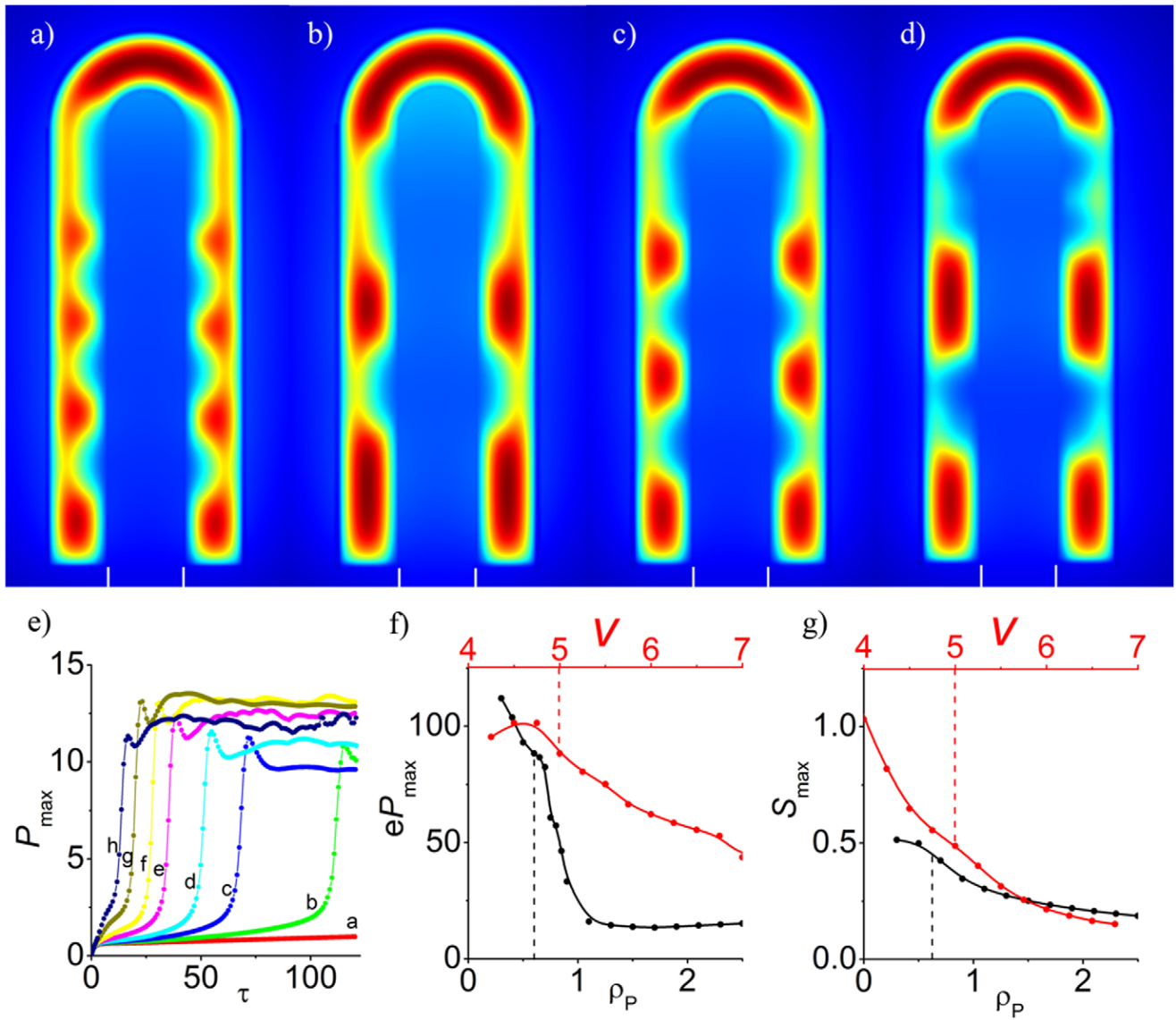

**Figure 6. Mutant phenotypes.** The FGF10 steady state concentration pattern in (**a**) wild-type (parameters as in Table 1), or (**b**) when the FGF expression rate was reduced to 75%, or the relative SHH diffusion constant increased by (**c**) 2-fold ($D_S = 10$) or (**d**) 6-fold ($D_S = 30$). Note the different spacings between FGF10 regions in the mutants. (**e**) The emergence of Ptc is delayed as the rate of SHH expression is reduced, i.e. $\rho_S = 200$ a, 300 b, 400 c, 500 d, 700 e, 900 f, 1300 g, 1700 h. (**f**) The effective rate of Ptc production decreases as the maximal rate of Ptc expression ($\rho_P$ and $v$) increases; this counterintuitive result is in agreement with experimental data. (**g**) The maximum SHH concentration decreases as the Ptc production rates $\rho_P$ and $v$ are increased. All parameter values are as in Table 1 unless otherwise stated.
doi:10.1371/journal.pcbi.1002377.g006

**Ptc-1.** While a mouse with lung specific over-expression of Ptc has not been created, other experiments show that increased expression of Ptc results in a reduction of SHH signaling, consequentially down-regulating expression of SHH responsive genes such as gli1 and Ptc itself [72]. The model reproduces this at first sight counterintuitive result and indeed predicts that the maximum concentration level of SHH as well as the rate of Ptc production decrease when the maximal Ptc expression rates ($\rho_P$ and $v$) are increased (Figure 6 f and g). Enhanced expression of Ptch1 expression in explants that are treated with FGF9 correlates with less epithelial buds (branches) [80]. We indeed observe less spots in our simulations when Ptch1 expression ($\rho_P$) is increased (data not shown).

## Discussion

We have developed a model of the core signaling interactions that governs lung branching morphogenesis (Figure 1a). We find that the experimentally described interactions give rise to a Schnakenberg-type Turing patterning mechanism which leads to the self-organized emergence of FGF10 pattern from homogenous initial conditions during *in silico* lung bud development. The predicted pattern are overall consistent with experimentally observed expression pattern both in wildtype and mutants, and the model predicts the reported increased spacing between FGF10 pattern in mutants with reduced FGF10 expression. The latter is a non-trivial prediction that cannot be explained with models reported earlier [36,42–44].

The parameter values are largely unknown for the developing lung, but many of the parameters have been established in other model systems (Table 2). To reduce the number of unknown parameters we non-dimensionalized the model and our conclusions thus only hinge on relative parameter values. In particular, we require that SHH and FGF10 diffuse about a 100-times faster than the receptor Ptc. Receptors are membrane proteins and are known to diffuse some 100–1000-fold more slowly than proteins in solution [55,56]. The diffusion of SHH and FGF10 in the fluid-filled cavities, $\bar{D}_{\text{ext}}$, is again some 10-fold higher as is common for unhindered diffusion of proteins [81]. Since the domain size and the time span of the developmental processes are well known also the absolute values of the parameters can be estimated from their non-dimensional counterparts (Table 2), and we note that all parameter values are within physiological ranges established in other model organisms, i.e. [52,53,81]. Finally we note that introducing the simplifying quasi-steadystate assumption for the formation of the SHH-Ptc complex does not affect the observed pattern as long as the non-dimensional binding and unbinding constants are of order 1 which is in the likely physiological range.

The model shows that the different branching modes could result from different patterns of FGF10. Thus for certain parameter ranges FGF10 accumulates at the tip of the lung bud. Since FGF10 induces outgrowth towards the highest FGF10 concentration this should lead to an elongation of the lung bud. As the lung bud is elongating at the tip new FGF10 spots emerge at the stalk of the lung bud close to the tip. These spots would lead to the lateral outgrowth of the lung bud and thus to lateral branching. For other parameter ranges FGF10 is absent from the lung tip and concentrates towards the sides. The lung bud would thus no longer elongate but grow out towards the sides which could be interpreted as the bifurcation branching mode. Since we only consider a 2D slice of the lung bud the model cannot differentiate between planar and orthogonal bifurcations.

Interestingly, most parameter values can alter the FGF10 distribution and thus the mode of branching (Figure 3). An increase in the diffusion constants or a lowering in domain size favour bifurcations while enhanced protein production or reduced decay tend to favour lateral branching. This may explain why genetically different individuals tend to have different branching patterns. At the same time this offers regulatory control to the many signaling factors that can affect these rates, but which we chose to ignore in this simple model that focuses on the core signaling proteins. Thus BMP signaling may lower the rate of FGF10 expression or interfere with the FGF-dependent increase in SHH expression. Both effects would favour bifurcations over lateral branching. BMP-dependent induction of Gremlin and Noggin on the other hand would lead to the sequestration of BMPs and thus limit the impact of BMPs. A strong increase or decrease in BMP activity may then move the SHH and FGF10 expression rates outside the patterning range. This may explain the detrimental impact of both Bmp overexpression and conditional knock-out on lung branching morphogenesis [32,33]. Further computational modelling combined with experimentation will be required to clarify this. The effective diffusion constants may be affected by the expression of glycoproteins as previously reported for the morphogen Dpp [82].

Growth has previously been reported to strongly affect pattern selection in Turing models [83]. Indeed the rate of growth also alters the branching pattern in our simulations (Figure 5). Thus we observe FGF10 pattern characteristic of lateral branching at high growth speeds and FGF10 pattern characteristic of bifurcations at low growth speeds. The growth rate $\bar{v}_g$ used to model the lateral branching mode in our model is around 14 $\mu m\ h^{-1}$ and gives rise to two new branches per day with branches separated by approximately 150–200 $\mu m$; this is well in agreement with experimental observations [17,24]. To simulate the bifurcation mode of branching we use a growth rate of 3.6 $\mu m\ h^{-1}$ which is close to the growth speed estimated from data in Metzger and co-workers [17].

We can speculate that the concentration of FGF10 affects the speed of outgrowth such that changes in the concentration during outgrowth would determine the sequence of the branching events. For our particular choice of parameters we observe lower FGF10 concentrations in spots close to the tips of rapidly growing lung buds (Figure 7) which may lead to bifurcations in the next generation. More proximal lateral spots have higher FGF10 concentrations, and these branches may grow out to longer length as indeed observed for some of the proximal branches. We further note that the position at which new branches appear depends on the growth mode. Thus if growth is at least 2–3 times faster at the

**Table 2.** Dimensional parameters.

| parameter | value | experimental range |
|---|---|---|
| $r_0, \mu m$ | 50 | $\approx 50$ [32,59] |
| $r_1, \mu m$ | 100 | $\approx 100$ [32,59] |
| $l_{ep}, \mu m$ | 10 | $\approx 10$ [32,59] |
| $\bar{v}_g, \mu m\ h^{-1}$ | 14 (3.6)[1] | $\approx 14(4)$ [17,24] |
| $\bar{D}_S, \mu m^2 s^{-1}$ | 12.5 | 0.1–50 [52,53] |
| $\bar{D}_F, \mu m^2 s^{-1}$ | 2.5 | 0.1–50 [52,53] |
| $\bar{D}_P, \mu m^2 s^{-1}$ | 0.05 | 0.001–0.5 [54–56] |
| $\bar{D}_{ext}, \mu m^2 s^{-1}$ | 100 | 10–200 [81] |
| $\bar{\delta}_F, s^{-1}$ | $5 \times 10^{-3}$ | $10^{-4} - 10^{-3}$ [52,53] |
| $\bar{\delta}_S, s^{-1}$ | $0.2 \times 10^{-3}$ | $10^{-4} - 10^{-3}$ [52,53] |
| $\bar{\delta}_P, s^{-1}$ | $1 \times 10^{-3}$ | $10^{-4} - 10^{-3}$ [52,53] |
| $\bar{\delta}_C, s^{-1}$ | $1.6 \times 10^{-3} \times C_1$ | - |
| $\bar{v}, s^{-1}$ | $5 \times 10^{-3} \times C_1$ | - |
| $\bar{\rho}_F, mol\ \mu m^{-3}\ s^{-1}$ | $3.5 \times 10^{-3} \times C_2$ | - |
| $\bar{\rho}_S, mol\ \mu m^{-3}\ s^{-1}$ | $1.6 \times C_3$ | - |
| $\bar{\rho}_P, mol\ \mu m^{-3}\ s^{-1}$ | $6 \times 10^{-4} \times C_3$ | - |
| [F], $mol\ \mu m^{-3}$ | $0.3 \times C_2$ | typical concentrations |
| [S], $mol\ \mu m^{-3}$ | $0.5 \times C_3$ | in the simulations |
| [P], $mol\ \mu m^{-3}$ | $10 \times C_3$ | in the region of expression |

[1]value without/with brackets corresponds to lateral branching/bifurcation modes of branching, respectively.
Note that the pattern-forming mechanism is highly robust to the exact protein concentrations as long as the relative ratios are preserved. The concentration-dependent parameters can take any value (as long as the relative ratios are preserved) depending on the parameters $C_1 = \bar{K}_F^{-2/3} \times \bar{\Gamma}^{-1/3} [mol\ \mu m^{-3}]$, $C_2 = \bar{K}_S [mol\ \mu m^{-3}]$, $C_3 = \bar{K}_F^{1/3} \times \bar{\Gamma}^{-1/3} [mol\ \mu m^{-3}]$ where $\bar{\Gamma} = \frac{\bar{k}_{on}}{\bar{k}_{off} + \bar{\delta}_C}$. With $\frac{\bar{r}_0^2}{\bar{D}_F} = 1000$ seconds and $r_0 = 50 \mu m$ the simulation time and length scales correspond to those observed experimentally [17,24], and we obtain this set of dimensional parameters.
doi:10.1371/journal.pcbi.1002377.t002

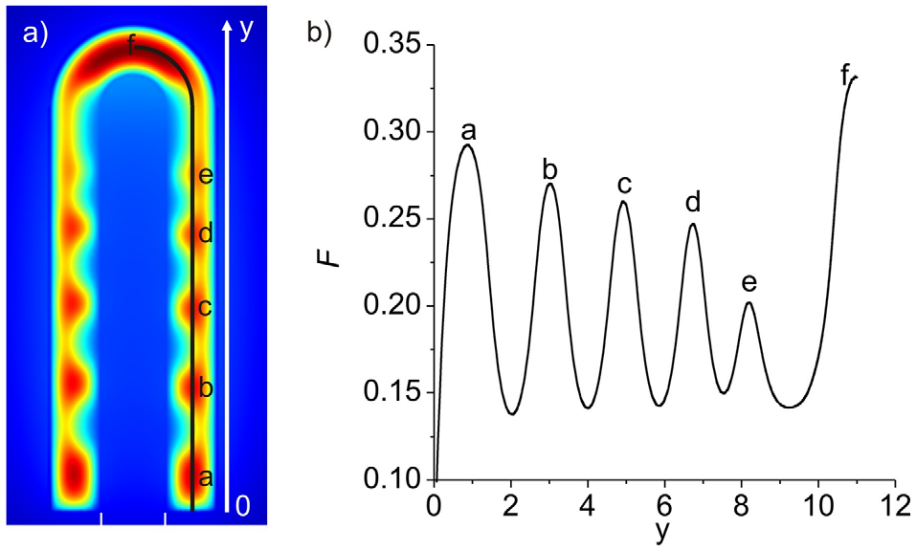

**Figure 7. The FGF10 concentration profile along the growing lung bud. a)** The FGF10 distribution in a rapidly growing lung bud ($v_g = 0.08$) at $\tau = 120$. The letters denote the different FGF10 concentration peaks; the profile is shown in panel b. **b)** The FGF10 concentration profile along the line shown in panel a. The labels in both panels denote the different branching points along the y-axis; peak f refers to the FGF10 concentration (F) in the lung tip. The FGF10 concentration is highest at the lung tip (f) and is lowest at the newly formed branched point behind the tip (e). The parameters used are as indicated in Table 1.
doi:10.1371/journal.pcbi.1002377.g007

tip then new FGF10 spots emerge directly behind the tip. In case of more uniform growth FGF10 spots appear also in the more proximal domain. Both growth modes and patterning dynamics have been observed during lung branching morphogenesis, and we therefore suggest that both growth modes exist in the developing lung and lead to the different patterning sequences.

Further advancements in our understanding of lung branching morphogenesis will require the development of three-dimensional models where the local growth rate is coupled to the FGF10 concentration and the inclusion of more signaling factors, most importantly those of BMPs. The parameterization and validation of such models will require new experimental data that reveal the dynamics of the three dimensional dynamics of branching and that quantify the epithelial and mesenchymal responses as well as the growth speed relative to the FGF10 concentration. Such information can now be acquired with the help of optical projection tomographs [84], and this method has been already used to capture the three dimensional dynamics of developing kidney [85]. Further advances in experimental techniques can thus be expected to provide exciting new insights into the regulatory processes of branching morphogenesis during organ development.

## Supporting Information

**Figure S1** FGF distribution at the steady state calculated according to the model presented by Hirashima *et al* [44]. SHH concentration is fixed at the lung tip and FGF production is promoted by SHH. The computational domain is equal to that shown on sub-figures a) and b). a) Elongation mode is observed when lung tip is far from the impermeable domain boundary, b) planar bifurcation is observed when the lung tip is in the proximity of the impermeable domain boundary. c) FGF expression pattern calculated on the infinitely big domain, in this case FGF distribution is always corresponds to elongation mode. Note, that panel c) is scaled differently compared to a) and b); stalk and tip radius are the same in all panels.
(TIF)

**Figure S2** Turing pattern on the continuos domain and on the array of cells. The steady state distribution of the fast diffusion component a), c) and slowly diffusion component b), d) in a Schnakenberg model. Upper and lower rows show the solution of the Schnakenberg model on 2D plane and on an array of spheres, correspondingly.
(TIF)

**Figure S3** The steady state distributions of the concentrations of FGF10, SHH and the receptor Ptc on a domain that is divided into cells. a) If the diffusion coefficient of Ptc at the cell edge is significantly lower than on the cell'' surface ($D_P = 0.001$) then the observed patterns are the same as in a continues model (compare to Figure 2a). b, c) The diffusion coefficient of Ptc is set to zero at the cell'' edge (black lines on the domain), panels b) and c) show pattern of the domain split into deformed rectangular and rectangular "cells", correspondingly. In this case the observed patterns are distorted from those observed in the case of the continuous model (Figure 2); however, all main features are preserved, in particular at the mesenchyme/epithelium border (right column). Unless stated otherwise the parameter values in Table 1 were used.
(TIF)

**Figure S4** The steady state distribution of FGF10. a) n = 1, b) n = 3 parameters of other values are similar to that presented in Table 1.
(TIF)

**Figure S5** The steady state distribution of FGF10 calculated assuming various SHH-Ptc complex stoichiometry a) $SP_3$, b) $S_2P_2$, c) $SP_{1.2}$. The values of the other parameters are similar to the ones in Table 1, except for case c) where Dp = 0.004.
(TIF)

**Figure S6** The impact of mesh resolution. FGF distribution at $\tau = 90$ calculated on a mesh with a maximum element size of a) 0.4, b) 0.2, c) 0.1 and d) 0.05.
(TIF)

**Figure S7 SHH, FGF and receptor Ptc expression patterns at the steady state.** Expression patterns of (**a,d**) FGF10, (**b,e**) SHH, and (**c,f**) Ptc in the steady state for parameter values as in Table 1 (**a–c**) or with $\rho_S = 1300$ (**d–f**). The upper panel presents an example of FGF10 distribution during the lateral branching mode, while the lower panel provides an example for FGF10 distribution during a bifurcation branching mode.
(TIF)

**Figure S8 FGF pattern in 3D.** The steady state distribution of FGF10 in 3D. Upper and lower panel show lateral branching and bifurcation modes of branching, correspondingly. Parameter values are as similar to those indicated in Table 1.
(TIF)

**Figure S9 The impact of growth mode on a pattern.** The FGF distribution on a growing lung: uniform growth a), local growth at the tip b). Parameters values used to simulate domain growth in the local growth mode are equal to that given in Table 1, except initial stalk length $h_0 = 0.2$ and in the case of domain stretching $\delta_S = 0.2$, $\delta_P = 1$, $\delta_C = 1.7$, $v = 5.1$, $\rho_F = 3$, $h_0 = 0.2$ and the rest of parameters are as given in Table 1.
(TIF)

**Figure S10 The robustness of FGF pattern to parameter variability.** FGF pattern calculated on a growing domain with parameter distributed normally with standard deviation equal to 0.1 a), 0.2 b) and 0.3 c) of the corresponding value given in Table 1.
(TIF)

**Figure S11 Dependence of FGF pattern in the wild type and FGF mutant lung on the lung growth rate.** The FGF pattern were simulated for (**top row**) wildtype conditions (Table 1) and (**bottom row**) mutants with 25% lower FGF expression ($\rho_F = 2.6$) on constant (SS) and growing domains (growth speed $vg$). Pattern are compared on equally sized domains.
(TIF)

**Figure S12 The steady state distributions of FGF10, SHH and receptor Ptc concentrations in a mesenchyme free lung in an a gel.** Panels a) and b) show patterns at different FGF10 concentrations.
(TIF)

**Table S1** Relation between dimensionless and dimensional parameters.
(EPS)

## Acknowledgments

The authors are grateful to Ross Metzger and Xin Sun for discussions and to Philipp Germann for Figure 1A, his help with implementing domain growth, and for critical reading of the manuscript.

## Author Contributions

Conceived and designed the experiments: DI CK DM. Performed the experiments: DM. Analyzed the data: DM. Contributed reagents/materials/analysis tools: DM CK DI. Wrote the paper: DI DM.

<sb><p>Branch Mode Selection during Lung Development</p></sb>